\documentclass[useAMS,usenatbib]{mn2e}
\usepackage{graphics,epsfig,aas_macros}
\usepackage{graphicx}
\usepackage[normalem]{ulem}
\usepackage[dvipsnames]{color}
\usepackage[]{inputenc,amssymb}
\usepackage{amsmath}
\usepackage{color}
\usepackage{textcomp}
\usepackage{array}
\usepackage{textcase}
\usepackage{titlesec}
\usepackage{etoolbox}
\usepackage{comment}
\usepackage{caption}
\usepackage[justification=centering]{caption}

\makeatletter
\patchcmd{\ttlh@hang}{\parindent\z@}{\parindent\z@\leavevmode}{}{}
\patchcmd{\ttlh@hang}{\noindent}{}{}{}
\makeatother
\renewcommand\footnoterule{\kern 5pt \hrule width 2in \kern 5.0pt}

\titleformat{\paragraph}
{\it\normalsize\normalsize}{\theparagraph}{1em}{}
\titlespacing*{\paragraph}
{0pt}{3.25ex plus 1ex minus .2ex}{1.5ex plus .2ex}

\newcolumntype{P}[1]{>{\centering\arraybackslash}p{#1}}

\usepackage[title]{appendix}

\def \be{\begin{equation}}
\def \ee{\end{equation}}
\def \bea{\begin{eqnarray}}
\def \eea{\end{eqnarray}}

\def \ms{{$M_*$} }
\def \mb{{$M_{\rm b}$}}
\def \mg{{$M_{\rm gas}$} }
\def \m5c{{$M_{500c}$} }

\def \mg{{$M_{\rm gas}$} }
\def \sigv{{$\sigma_v$} }
\def \tg{{$T_{\rm gas}$} }
\def \mv{{$M_{\rm vir}$} }
\def \yx{{$Y$} }
\def \lx{{$L_{\rm bol}$} }
\def \om{{$\Omega_m$} }
\def \ob{{$\Omega_b$} }

\title[Cosmological dependence in MOR]{Cosmology dependence of galaxy cluster scaling relations}

\author [P. Singh et al.]
{Priyanka Singh$^{1,2}$ \thanks{priyanka.singh@inaf.it}, 
	Alex Saro$^{1,2,3}$, Matteo Costanzi$^{1,2}$, Klaus Dolag$^{4,5}$
	\\\\
	$^1$ INAF-Osservatorio Astronomico di Trieste, via G. B. Tiepolo 11, I-34143 Trieste, Italy\\
    $^2$ IFPU - Institute for Fundamental Physics of the Universe, Via Beirut 2, 34014 Trieste, Italy\\	
    $^3$ Astronomy Unit, Department of Physics, University of Trieste, via Tiepolo 11, I-34131 Trieste, Italy\\
    $^4$ University Observatory Munich, Scheinerstr 1, D-81679 Munich, Germany\\
    $^5$ Max-Planck-Institut f\"{u}r  Astrophysik (MPA), Karl-Schwarzschild Strasse 1, D-85748 Garching bei M\"{u}nchen, Germany\\
}

\begin{document}
	\label{firstpage}
	% \pagerange{\pageref{firstpage}--\pageref{lastpage}}
	\maketitle
	
	\begin{abstract}
	    The abundance of galaxy clusters as a function of mass and redshift is a well known powerful cosmological probe, which relies on underlying modelling assumptions on the mass-observable relations (MOR). Some of the MOR parameters can be constrained directly from multi-wavelength observations, as the normalization at some reference cosmology, the mass-slope, the redshift evolution and the intrinsic scatter. However, the cosmology dependence of MORs cannot be tested with multi-wavelength observations alone. We use {\tt Magneticum} simulations to explore the cosmology dependence of galaxy cluster scaling relations. We run fifteen hydro-dynamical cosmological simulations varying \om, \ob, $h_0$ and $\sigma_8$ (around a reference cosmological model). The MORs considered are gas mass, baryonic mass, gas temperature, \yx and velocity dispersion as a function of virial mass. We verify that the mass and redshift slopes and the intrinsic scatter of the MORs are nearly independent of cosmology with variations significantly smaller than current observational uncertainties. We show that the gas mass and baryonic mass sensitively depends only on the baryon fraction, velocity dispersion and gas temperature on $h_0$, and \yx on both baryon fraction and $h_0$. We investigate the cosmological implications of our MOR parameterization on a mock catalog created for an idealized eROSITA-like experiment. We show that our parametrization introduces a strong degeneracy between the cosmological parameters and the normalization of the MOR. Finally, the parameter constraints derived at different overdensity ($\Delta_{500c}$), for X-ray bolometric gas luminosity, and for different subgrid physics prescriptions are shown in the appendix.
	\end{abstract}
	
	
\section{Introduction}
Galaxy clusters are the most massive gravitationally bound structures in the Universe and represent a well known powerful cosmological tool. Their abundance as a function of redshift and mass is sensitive to both the expansion history and the history of structure formation in the Universe (see \citealt{allen11} for a review), providing therefore complementary information to purely geometric probes such as Type Ia supernovae, the primary cosmic microwave background (CMB) and baryonic acoustic oscillations.

Integrated observable properties of galaxy clusters like X-ray luminosity and temperature, the optical richness and their associated velocity dispersion, and the intensity of the Sunyaev-Zeldovich effect (SZE: \citealt{sunyaev72}), are generally used as a proxy for the total cluster mass, as they are expected to regularly scale with galaxy cluster mass following mass-observable scaling relations (MOR), although with some associated intrinsic scatter. Current studies of the cluster mass function (often described as cluster number-counts experiments) are therefore simultaneously exploring both cosmological and MOR (including the intrinsic scatter) parameters to constrain cosmological models \cite[e.g.,][]{planck15cluscosm,mantz15,bocquet19}. 
The standard approach for these state-of-the-art studies is to calibrate the MOR empirically, by anchoring the associated parameters %through either weak-lensing \citep[e.g.][]{bardeau07, okabe10, hoekstra12, marrone12, applegate14, umetsu14, gruen14, hoekstra15, okabe16, battaglia16, applegate16, hilton18}, or through dynamical studies \citep[e.g.][]{sifon12, hasselfield13, bocquet15, capasso19a, capasso19b, capasso19c}, 
 through either weak-lensing \citep[e.g.][]{bardeau07, okabe10, hoekstra12, applegate14, hoekstra15, hilton18, dietrich19} or through dynamical studies \citep[e.g.][]{sifon12, hasselfield13, bocquet15, capasso19a, capasso19b, capasso19c},
methods which are more directly linked to effect of gravity alone, and thus easy to characterize in terms of systematics associated with the treatment of the complex physics regulating the baryonic component. Therefore, for this reason, biases associated with weak lensing and dynamical estimates can be calibrated more robustly with numerical simulations. As a result, MOR parameters including the normalization, the mass slope, the redshift evolution, and the scatter, can be directly constrained from multi-wavelength observations \citep{mantz15, dietrich19, bocquet19}. In most of these studies, the cosmological dependence of MORs is usually assumed to be only related to the background evolution of the Universe, with the notable exception of measurements of the baryon fraction in galaxy cluster to constraint the matter density \citep[e.g.,][]{mantz14}. 

Within this framework, numerical and hydro-dynamical cosmological simulations still provide fundamental information:
\vspace{-0.5\baselineskip}
\begin{enumerate}
    \item they provide accurate calibration of the theoretical halo mass function \citep[e.g.,][and references therein]{tinker08, cui14,velliscig14,bocquet15,despali16,mcclintock19};
    \item they provide accurate calibration of possible biases affecting the observables used to anchor the absolute scale of the MOR, as weak-lensing and dynamical mass calibration \citep[e.g.,][and references therein]{white10,saro13,becker11,rasia12};
    \item they provide guidance on the functional form for the mean relation and associated intrinsic scatter of the MOR, as well as priors on parameters that observationally are on weakly constraint \citep{stanek09,truong18,gupta16}. 
\end{enumerate}
In particular, as we have access to only one observable Universe, these simulations represent the only way to test if the MORs are cosmology dependent and, if they are, to calibrate the parameters describing this dependence. 

In this paper, we explore the cosmology dependence of galaxy cluster scaling relations using cluster catalogs identified in the suite of {\tt Magneticum} simulations\footnote{www.magneticum.org}. These large cosmological simulations are designed to investigate different cosmological scales with very large number of particles and, at the same time, to describe the hydro-dynamical evolution of the baryonic collisional component. They therefore provide a complementary tool with respect to purely N-body simulations such as {\tt Quijote} simulations \citep{navarro19}, the {\tt Mira-Titan Universe} \citep{heitmann16, lawrence17} and {\tt Aemulus} simulations \citep{derose19}. Furthermore, due to the large simulated cosmological volumes, they are better designed to study the cluster population with respect to other higher-resolution, but smaller volumes simulations such as, e.g., the {\tt BAHAMAS} simulations \citep{mccarthy18, mccarthy19}. Moreover, the purpose of {\tt Magneticum} simulations is to provide a theoretical counterpart for Large Scale Structure (LSS), therefore the dynamical range of the cosmological parameters space explored is significantly broader than the current CMB cosmological constraints \citep{planck17}, currently tested in other studies (e.g. {\tt Aemulus}, {\tt BAHAMAS} simulations). 

%They are also aimed to provide powerful, realistic mocks for many existing and upcoming large sky multi-wavelength surveys (eg. \citealt{dolag16, soergel17, biffi2018}).

In this work, we run {\tt Magneticum} simulation for fifteen different cosmological volumes, each one generated with the same initial seeds, but with different cosmological parameters. All the simulations include the description of the same physical processes and use the same sub-grid model parameters. Our basic assumption is that the variation of the cosmological model should not directly affect the microscopic processes that these sub-grid parameters describe and therefore, these sub-grid physics parameters have been tuned to reproduce observed properties of galaxy clusters at an arbitrarily (but reasonable, and consistent with observations) chosen reference cosmology \citep{bocquet16, gupta16, dolag17, remus17, biffi2018, ragagnin19}. In other words, their numerical value does not carry any physical meaning. 
As a result, the results presented in this work could be considered robust only if they are independent of the (arbitrary) choice of the reference cosmology used to tune the sub-grid model parameters. 
Instead of re-tuning them for each different cosmological model (an effort which would require an unfeasible computational cost), we address the robustness of our results by studying the  validity of our model assumption in the adopted functional form of the MORs. 
In particular, we explicitly verify that the cosmological dependence of the MORs is only affecting the normalization of the studied scaling relations, but does not change the mass slope, the redshift evolution, and the intrinsic scatter. Therefore, we argue that re-tuning the subgrid model parameters for different choices of the reference cosmology would not impact the cosmological dependence of the MOR, as it will only translate into a different zero-point normalization. 
 
 %Moreover, tuning the subgrid physics at every cosmology may introduce large degeneracies between cosmological and subgrid model parameters. It may also force us to an unrealistic subgrid model to compensate for a wrong cosmology while trying to match the observations. A detailed examination of degeneracies between cosmology and subgrid physics is beyond the scope of this paper. {\ccr what do you think of this approach? or shall we also talk about the extra subgrid models in the intro?}

In summary, the aim of this paper is: $i)$ to construct a universal scaling relation where the cosmology dependence of the scaling relation is absorbed in its amplitude, $ii)$ to test the robustness of our parameterization with respect to the observational uncertainties, and $iii)$ to forecast the impact our cosmology dependent parameterization on an idealized cluster cosmology experiment. This paper is organized as follows: in Section \ref{sec-sim} we briefly describe the details of the simulation setup. In Section \ref{sec-method}, we describe the basic ingredients of the MOR and test its robustness. In Section \ref{sec-results}, we present the results of our analysis for
$M_{\rm gas}$, $T_{\rm gas}$, $Y$ and $\sigma_v$. In Section \ref{sec-impact}, we discuss the impact of our MOR parametrization for an idealized cosmology experiment. In Section \ref{sec-summary}, we present the summary of the main analysis.

	\begin{table*}
	\centering
	\caption{Cosmological parameter values for the fifteen simulation boxes.}
	\resizebox{1.0\textwidth}{!}{
		\setlength{\tabcolsep}{3pt}
		\begin{tabular}{c | c c c c c c c c c c c c c c c}
			
			& C1 & C2 & C3 & C4 & C5 & C6 & C7 & C8 & C9 & C10 & C11 & C12 & C13 & C14 & C15\\\\
			\hline 
			\\
			\om & 0.153 & 0.189 & 0.200 & 0.204 & 0.222 & 0.232 & 0.268 & 0.272 & 0.301 & 0.304 & 0.342 & 0.363 & 0.400 & 0.406 & 0.428 \\\\
			\ob & 0.0408 & 0.0455 & 0.0415 & 0.0437 & 0.0421 & 0.413 & 0.0449 & 0.0456 & 0.0460 & 0.0504 & 0.0462 & 0.0490 & 0.0485 & 0.0466 & 0.0492\\\\
			$\sigma_8$ & 0.614 & 0.697  & 0.850 & 0.739 & 0.793 & 0.687 & 0.721 & 0.809 & 0.824 & 0.886 & 0.834 & 0.884 & 0.650 & 0.867 & 0.830 \\\\
			$h_0$ & 0.666 & 0.703 & 0.730 & 0.689 & 0.676 & 0.670 & 0.699 & 0.704 & 0.707 & 0.740 & 0.708 & 0.729 & 0.675 & 0.712 & 0.732 \\\\
			$f_b$ & 0.267 & 0.241 & 0.208 & 0.214 & 0.190 & 0.178 & 0.168 & 0.168 & 0.153 & 0.166 & 0.135 & 0.135 & 0.121 & 0.115 & 0.115 \\	
			%	\hline
		\end{tabular}
		\label{tab-cosmo}}
\end{table*}

\section{Simulation details}
\label{sec-sim}
{\tt Magneticum} simulations are based on the Smoothed Particle Hydrodynamics (SPH) code {\tt P-GADGET3} which itself is an improved version of {\tt P-GADGET2} \citep{springel05a, springel05b}. The simulation includes a variety of physical processes such as metallicity dependent radiative cooling \citep{wiersma09}, UV background heating \citep{Haardt01}, a detailed model of star formation, chemical enrichment \citep{tornatore07} and supernovae (SNe) as well as active galactic nuclei (AGN) driven feedback prescriptions \citep{springel03, dimatteo08, fabian10, hirschmann14, bocquet16}.
	
For the purpose of our study, we use Box1a from {\tt Magneticum} simulation set, which is a large size, medium resolution box. The size of the box is $\sim$ 896 $h_0 ^{-1}$Mpc. It contains $1526^3$ dark matter particles and an equal number of gas particles. For our reference cosmology, this corresponds to a characteristic mass resolution of dark matter, gas and star particles of $1.3\times 10^{10}$ $h_0 ^{-1} M_{\odot}$, $2.6 \times 10^{9}$ $h_0 ^{-1} M_{\odot}$ and $6.5\times 10^{8}$ $h_0 ^{-1} M_{\odot}$, respectively. The gravitational softening lengths for dark matter, gas and star particles are 10$h_0 ^{-1}$kpc, 10$h_0 ^{-1}$kpc and 5$h_0 ^{-1}$kpc, respectively.

\begin{figure*}
		\includegraphics[width=14.0cm,angle=0.0]{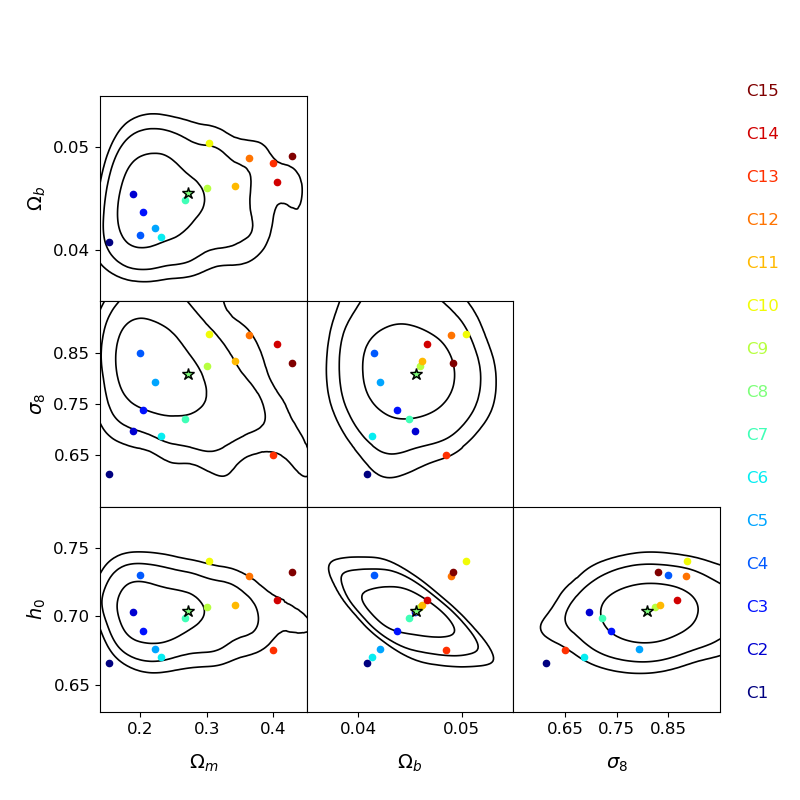}
		\caption{Distribution of cosmological parameters for the fifteen cosmologies used in this paper. WMAP7 (i.e. our C8) is highlighted by star symbols. The black contours represent 68\%, 95\% and 99\% confidence limits on these parameters \citep{bocquet19} with additional Gaussian priors applied on $h_0$.}
		\label{fig-cosmo}
	\end{figure*}

We run the same simulation set-up for a sample of fifteen different flat $\Lambda$CDM cosmological models (C1, C2,..., C15). The cosmological parameters varied in each simulations are $\Omega_m$, $\sigma_8$, $h_0$, and $\Omega_b$ and their values are specified in Table \ref{tab-cosmo} and shown in Figure \ref{fig-cosmo} as coloured points, together with cosmological constraints obtained by state-of-the-art cluster number counts experiment \citep[][with an additional Gaussian prior on $h_0$ with mean 0.704 and width 0.014]{bocquet19} . The parameter ranges used are thus $0.15 < \Omega_m < 0.45$, $0.6 < \sigma_8 < 0.9$, and $0.65 < h_0 < 0.75$, to cover the entire dynamic range of current large-scale-structure cosmological constraints. The cosmological parameters are chosen from above ranges using Latin hypercube sampling\footnote{https://pythonhosted.org/pyDOE/randomized.html}. The cosmologies are labelled as C1, C2,...., C15 in order of increasing value of $\Omega_m$. Note that, thirteen out of fifteen cosmologies have fixed $\Omega_b h^{-2} _0 \sim 0.092$, perpendicular to the direction of degeneracy between \ob and $h_0$. The cosmologies C3 and C13 have been added to break the degeneracy between the two. Our reference cosmology (C8) corresponds to WMAP7 best fit results \citep{komatsu11}.

Friends-of-friends (FoF) algorithm is used to identify haloes, linking only the dark matter particles with a linking length b = 0.16. A {\tt SUBFIND} algorithm \citep{springel01, dolag09} is used to compute spherical overdensity (SO) virial mass \citep[Mvir,][]{bryan98}, where, $\Delta_{\rm vir}$ is a function of $C_i$. The halo is centred at the position of the dark matter particle in a FoF group having the minimum value of the gravitational potential. The observable quantities are integrated within the virial radius for each of the identified halo. Specifically, we study the gas mass (\mg), the gas temperature ($T_{\rm gas}$), the X-ray pseudo-pressure $Y$ ($\equiv$ \mg $\times$ \tg), where \mg is the sum of mass of all gas particles within a given overdensity radius, and \tg is the associated mass weighted gas temperature. Furthermore we also investigate the mass-$\sigma_v$ relation, where the velocity dispersion $\sigma_v$ is the mass weighted velocity dispersion of all the particles belonging to each main halo.
Results for X-ray bolometric gas luminosity and stellar mass are given in Appendix \ref{sec-lx} and \ref{sec-ms}, respectively. Results for \m5c ($\Delta_{500c}=500$ w.r.t. $\rho_c$) are given in Appendix \ref{sec-m500c}.

We apply conservative mass cuts and select only haloes with $M_{\mathrm{vir}} > 2\times 10^{14} M_\odot$ (corresponding to approximately 10$^4$ particles), to ensure that haloes extracted from the hydro simulations are not affected by issues related to resolution and numerical artifacts. We extract cluster catalogues at six redshifts. The final number of haloes used in each simulation and redshift is shown in Table \ref{tab-numofclus}.
    
We note that all the above mentioned observables are representing simplified and idealized versions of the actual physical observed quantities. A large effort has been dedicated over the last decades (e.g., \citealt{nagai07, avestruz14}) to understand how the physical properties of haloes translate into observables (and vice-versa) through the analysis of dedicated mocks. These mock simulation works quantifies observational biases and scatter in observationally derived quantities, such as gas mass \citep{nagai11, zhuravleva13}, X-ray temperature, $Y$ \citep{khedekar13, rasia14}, and velocity dispersion \citep{lau11,saro13, munari13}, and hydrostatic mass \citep{lau09, lau13, nelson14, biffi16, shi16}. This project is however focused only on the cosmological dependence of the MORs, which we thus assume to be unrelated to the biases and scatter associated to such observational effects.
 
	\section{Method}
	\label{sec-method}
	For a given observable $O$, our aim is to construct a universal scaling relation, described by a set of parameters which are therefore assumed to be all mass, redshift and cosmology independent.  We adopt the following functional form to describe the MOR:
	%which incorporates the mass, redshift and cosmology dependence of various galaxy cluster observables. The universal scaling relation is given by,
	
	\begin{align}
	\ln O = \Pi_c + (\alpha+\alpha_{ss}) \ln\Big( \frac{M}{M_P}\Big) + \beta_{ss} \ln\Big( \frac{F(z)}{F(z_P)}\Big) \nonumber \\ 
	+ \beta \ln\Big( \frac{1+z}{1+z_P}\Big) \pm \sigma,
	\label{eqn-obs}
	\end{align}
	where $F(z) \equiv E(z) \sqrt {\Delta_{\mathrm vir}(z)}$, $\sigma$ is the intrinsic log-normal scatter, $\alpha_{ss}$ and $\beta_{ss}$ (see Table \ref{tab-ss}) represent the self-similar mass and
	redshift dependence of the MOR \citep{bohringer12}, and  $M_P = 2.85\times 10^{14} M_{\odot}$ and $z_P = 0.14$ are the pivot mass and redshift (the
	median mass and redshift of the sample). The parameters $\alpha$ and $\beta$ capture any deviations from the predicted self-similar mass dependence and evolution, and are therefore
	both zero in a perfectly self-similar scenario. We choose a simple 1+z parameterization to capture the redshift evolution of the MORs, separating it from the traditional $E(z)$ evolution. This way, we remove the expected cosmology dependent part (the E(z) term) associated to the self-similar evolution and be agnostic on the remaining deviations.
	The term $\Pi_c$ is the normalization of the MOR containing its cosmological dependencies. We assume a simple power law dependence of normalization on cosmological
	parameters.
	
	\begin{table*}
	\caption{The final number of haloes used in our analysis at different snapshots in different cosmological boxes above the lower mass limit $M_\mathrm{vir} > 2\times 10^{14} M_\odot$ (corresponding to approximately 10$^4$ particles).}
	\centering
	\resizebox{1.0 \textwidth}{!}{
		\setlength{\tabcolsep}{6pt}
		\begin{tabular}{c c c c c c c c c c c c c c c c}
			\\
			z & C1 & C2 & C3 & C4 & C5 & C6 & C7 & C8 & C9 & C10 & C11 & C12 & C13 & C14 & C15 \\\\
			\hline \\
             0 & 487 & 1698 & 4639 & 2945 & 5063 & 2998 & 4864 & 7823 & 10454 & 12293 & 14103 & 17817 & 8711 & 21879 & 20912 \\\\
             0.14 & 350 & 1335 & 3847 & 2299 & 4065 & 2262 & 3677 & 6150 & 8386 & 10093 & 11340 & 14570 & 6003 & 17818 & 16701 \\\\
             0.29 & 219 & 910 & 2976 & 1641 & 3033 & 1465 & 2547 & 4601 & 6116 & 7729 & 8413 & 11172 & 3644 & 13550 & 12258 \\\\
             0.47 & 121 & 516 & 2050 & 1034 & 2015 & 857 & 1460 & 3059 & 4102 & 5290 & 5607 & 7742 & 1861 & 9369 & 8087 \\\\
             0.67 & 59 & 254 & 1271 & 553 & 1144 & 396 & 711 & 1694 & 2351 & 3297 & 3244 & 4744 & 754 & 5577 & 4576 \\\\
             0.90 & 15 & 103 & 645 & 218 & 559 & 141 & 272 & 790 & 1091 & 1673 & 1558 & 2436 & 240 & 2872 & 2185 \\\\
			\hline
		\end{tabular}
		\label{tab-numofclus}}
\end{table*}

	\begin{equation}
	\Pi_c = \Pi_{c,0} + \gamma_{h_0}\ln \Big(\frac{h_0}{h^P _0}\Big) 
	+ \gamma_{b}\ln \Big(\frac{f_b}{f^P _b}\Big)
	+ \gamma_{\sigma_8}\ln \Big(\frac{\sigma_8}{\sigma^P _8}\Big) 
	\label{eqn-pic}
	\end{equation}
	where $f_b = \Omega_b/ \Omega_m$ is the cosmic baryon fraction, and $h_0^P = 0.704$, $f_b^P = 0.168$ and $\sigma_8^P = 0.809$ are the pivot points
	in cosmological parameters equal to our C8 cosmology. Note that, the cosmology dependence of $\Pi_c$ should be viewed as an "effective cosmology dependence" when the commonly used functional form of scaling relations is adopted. 
	
	The adopted MOR functional form (Equation \ref{eqn-obs} and \ref{eqn-pic}) consists therefore in two distinct families of parameters: 
	\begin{itemize}
	    \item Astrophysical parameters: $\Pi_{c,0}$, $\alpha$, $\beta$ and $\sigma$. These are the parameters that can be observationally constrained using multi-wavelength data.
	    \item Cosmological parameters $\gamma$s: $\gamma_{h_0}$, $\gamma_{b}$ and $\gamma_{\sigma_8}$. These are the parameters that cannot be constrained with observations. 
	\end{itemize}
	The primary purpose of this work is to provide the best estimates and most reliable uncertainties associated to the parameters describing the cosmology dependence of the MOR $\gamma$s. Note that, a running of $\Pi_c$ (i.e. a non-zero value of $\gamma$) does not imply a causation relation between the cosmological parameters and the scaling relation parameters. It incorporates any possible correlation between the MORs and the derived cosmology dependent quantities. 
	
\begin{figure*}
	\hspace*{-15mm}
	%	\centering
	\includegraphics[width=21.0cm,angle=0.0]{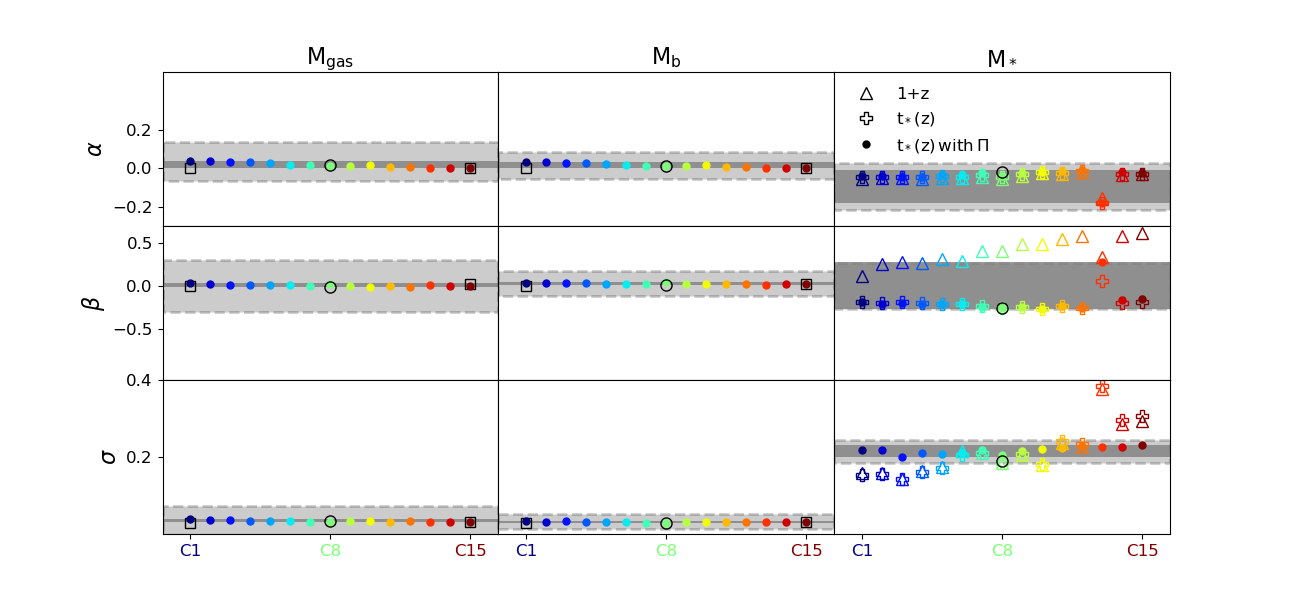}
	\caption{Variation in best-fitting values of $\alpha$, $\beta$ and $\sigma$ as a function of cosmology for \mg-\mv, \mb-\mv and \ms-\mv scaling relations. Filled circles correspond to the cosmologies shown in Figure \ref{fig-cosmo} with the same color scheme. The lighter grey bands enclosed by dashed lines represent current state-of-the-art observational uncertainties \citep{bulbul18, chiu2018}. Empty squares correspond to non-radiative runs (performed for C1 and C15) whereas empty circles correspond to the simulation run with a different feedback scheme for C8 cosmology. Darker grey bands represent the systematic uncertainty range quoted in Table \ref{tab-bf}.}
	\label{fig-uncertain-masses}
\end{figure*}

\begin{figure*}
	\hspace*{-15mm}
	%	\centering
	\includegraphics[width=21.0cm,angle=0.0]{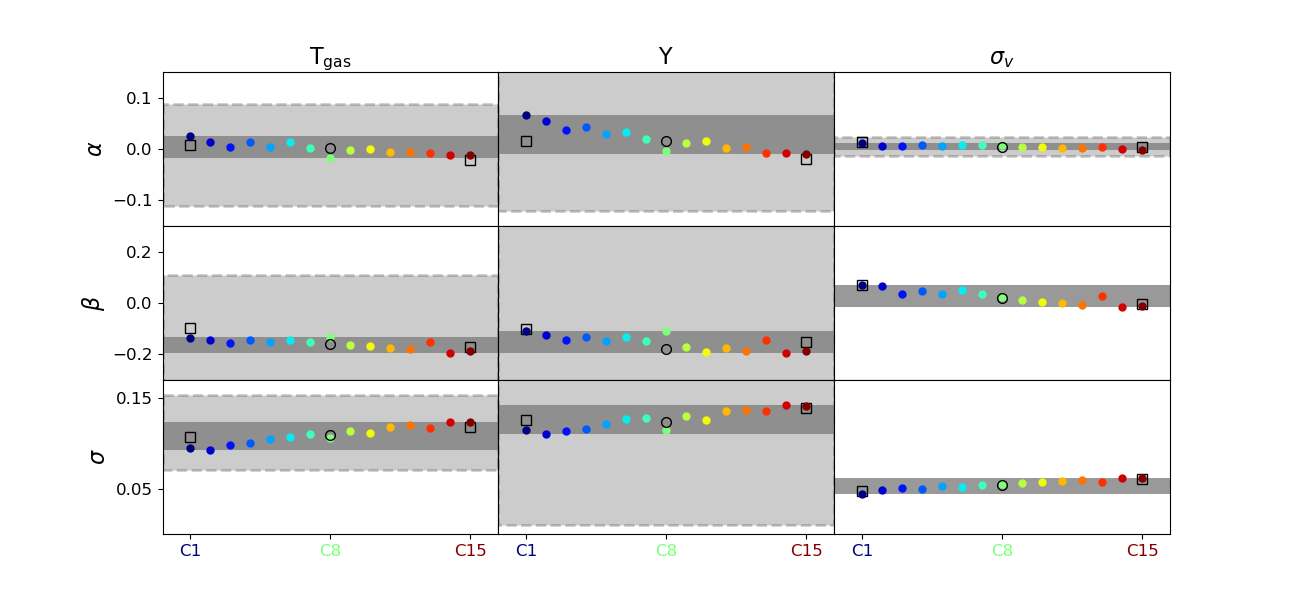}
	\caption{Same as Figure \ref{fig-uncertain-masses} for \tg-\mv, \yx-\mv and \sigv-\mv scaling relations. The uncertainty on $\sigma_v$-$M$ relation is the result of previous simulation studies \citep{evrard08}.}
	\label{fig-uncertain}
\end{figure*}

	\begin{table}
	\caption{Self-similar mass ($\alpha_{ss}$) and redshift ($\beta_{ss}$) dependence of the MORs.}
	\centering
	\resizebox{0.45 \textwidth}{!}{
		\setlength{\tabcolsep}{10pt}
		\begin{tabular}{c c c c c c c}
			\\
			& \mg & \mb & \ms & \tg & \yx & \sigv \\\\
			\hline \\
			$\alpha_{ss}$ & 1 & 1 & 1 & 2/3 & 5/3 & 1/3 \\\\
			$\beta_{ss}$ & 0 & 0 & 0 & 2/3 & 2/3 & 1/3 \\\\
			\hline
		\end{tabular}
		\label{tab-ss}}
\end{table}

    \subsection{Robustness of the model}
	\label{sec-unc}
	The main assumption that goes into constructing Equation \ref{eqn-obs} and \ref{eqn-pic} is that $\alpha$, $\beta$ and $\sigma$ are cosmology independent, or, 
	in other words, that cosmology is only affecting the normalization of the MOR. To test that our assumption is indeed justified, we fit the MOR (after applying a 3$\sigma$ clipping to remove outliers) separately on
	each individual cosmological simulation. More in detail, for each C$i$ simulation we use a Gaussian likelihood and uninformative priors  (with a support significantly broader than the recovered posterior probability distribution) and fit only for $\Pi_{c,0}$, $\alpha$, $\beta$ and $\sigma$ using Equation \ref{eqn-obs}:% to see whether $\alpha$, $\beta$ and $\sigma$ depend on C$i$:
	\begin{align}
	   \mathrm{ln }\mathcal{L} = -\frac{1}{2}\sum_j \Big[\Bigl(\frac{\ln O(\Pi_{c,0}, \alpha, \beta, \sigma, M_j, z_j)- \ln O_j}{\sigma}\Big)^2 \nonumber \\ + \ln(2\pi \sigma^2) \Big]
	\end{align}
	where $M_j$ and $z_j$ are respectively the mass and redshift of $j^{th}$ halo. The parameter space is explored with the {\tt emcee} affine-invariant sampler \citep{emcee} to find the best-fitting values and associated uncertainties.
%	Further details about the fitting procedure are presented in Section \ref{sec-systamatic}.

	Coloured points in Figures \ref{fig-uncertain-masses} and \ref{fig-uncertain} show the variation of $\alpha$, $\beta$ and $\sigma$ for the 15 different simulated cosmologies. The overall resulting range is highlighted by the darker grey bands.
	While residual trends and variations in these parameters are clearly visible (e.g., decreasing $\alpha$ for $M_\mathrm{gas}$ from C1 to C15), we 
	note that these differences are much smaller than current state-of-the-art observational constraints (lighter grey bands enclosed by dashed lines in Figures \ref{fig-uncertain-masses} and \ref{fig-uncertain}) for the three X-ray observables \cite{bulbul18} (hereafter, B19) and by previous studies on simulations for the $\sigma_v$-$M$ relation \citep{evrard08}. This result therefore justifies our assumption about the cosmology independence of $\alpha$, $\beta$ and $\sigma$ in Equation \ref{eqn-obs}. In other words, any cosmology dependence introduced in these parameters due to the presence of baryons is only of second order. Note that, the cosmology dependence of these astrophysical parameters cannot be neglected once it becomes comparable to the observational uncertainties. As next generation surveys will approach this limit, these approximations will be no longer acceptable and a much more careful description of the cosmological dependence of the MOR will be required, including wider parametrizations or relying on different approaches (e.g., emulators {\tt Quijote}, {\tt Mira-Titan Universe}, {\tt Aemulus}).
	
	In order to estimate how the residual lack of universality on the astrophysical parameters propagates into systematic variations on the cosmological parameters $\gamma$s, we then proceed as follows. 

    Instead of jointly fitting the astrophysical and cosmological parameters together for all the 15 C$i$ cosmologies, we fit only for the cosmological parameters $\gamma$s while keeping the astrophysical parameters fixed (except $\sigma$). We repeat the procedure 15 times, each time with the astrophysical parameters fixed to their best fitting values of each individual C$i$ cosmology. This procedure is therefore equivalent to weighting every cosmology equally. 
    For example, we fix the values of $\Pi_{c,0}$, $\alpha$ and $\beta$ to their best-fitting values for C1 and then run MCMC analysis to find best-fitting values of $\gamma$s. This analysis is repeated for all fifteen cosmologies, and therefore it gives us a set of fifteen best-fitting values of $\gamma_{h_0}$, $\gamma_{b}$ and $\gamma_{\sigma_8}$. The range of these cosmological parameters $\gamma$s exceeds the pure statistical uncertainty of the fit, which is driven by the overall extremely large number of objects in our simulations and therefore does not reflect the underlying limitations of our modelling. 
    The resulting range for both the astrophysical and cosmological MOR parameters are shown in Table \ref{tab-bf} and \ref{tab-bf-gammas}, respectively. The mean values quoted in these tables are simply the mean of the systematic uncertainty range.
   
	\section{Results}
	\label{sec-results}
	We present here our results for each of the four studied MORs. Note that, the observational results are generally at a different overdensity. Thus, the comparison of our results with the observations in this and the following sections is for qualitative purpose only.
	
	\begin{table*}
		\caption{Mean values and systematic uncertainties for astrophysical MOR parameters (described in detail in Section \ref{sec-unc}). The mean values quoted here are the mean of the systematic uncertainty range with symmetric error-bars.}
		\centering
		\resizebox{0.6\textwidth}{!}{
			\setlength{\tabcolsep}{6pt}
			\begin{tabular}{c c c c c c c}
				\\
				& \mg & \mb & \ms & \tg & \yx & \sigv \\\\
				\hline \\
				$\Pi_{c,0}$ & 31.4$\pm$0.4 & $31.45 \pm 0.39$ & 28.07 $\pm$ 1.41 & 0.5$\pm$0.2 & 31.9$\pm$0.2 & 6.4$\pm$0.1 \\\\
				$\alpha$ & 0.02$\pm$0.02 & $0.02 \pm 0.02$ & -0.09 $\pm$ 0.08 & 0.0$\pm$0.02 & 0.03$\pm$0.04 & 0.01$\pm$0.07 \\\\
				$\beta$ & 0.01$\pm$0.02 & $0.03 \pm 0.01$ & 0.003 $\pm$ 0.275 & -0.16$\pm$0.03 & -0.15$\pm$0.04 & 0.03$\pm$0.04 \\\\
				$\sigma$ & $<0.04$ & $< 0.03$ & 0.22 $\pm$ 0.02 & 0.11$\pm$0.02 & 0.13$\pm$0.02 & 0.05$\pm$0.01 \\\\
				\hline
			\end{tabular}
			\label{tab-bf}}
	\end{table*}

	\begin{table*}
	\caption{Same as Table \ref{tab-bf} for cosmological MOR parameters.}
	\centering
	\resizebox{0.6\textwidth}{!}{
		\setlength{\tabcolsep}{6pt}
		\begin{tabular}{c c c c c c c}
			\\
			& \mg & \mb & \ms & \tg & \yx & \sigv \\\\
			\hline \\
			$\gamma_{h_0}$ & -0.13$\pm$0.22 & $0.02 \pm 0.21$ & 1.20$\pm$0.55 & 0.78$\pm$0.05 & 0.65$\pm$0.20 & 0.38$\pm$0.04 \\\\
			$\gamma_{b}$ & 0.80$\pm$0.03 & $0.87 \pm 0.03$ & 2.49$\pm$0.1 & -0.02$\pm$0.01 & 0.78$\pm$0.03 & -0.05$\pm$0.01 \\\\
			$\gamma_{\sigma_8}$ & -0.14$\pm$0.06 & $-0.10 \pm 0.05$ & 1.71$\pm$0.28 & 0.14$\pm$0.04 & 0.01$\pm$0.05 & -0.01$\pm$0.01 \\\\
			\hline
		\end{tabular}
		\label{tab-bf-gammas}}
\end{table*}

	\subsection{\mg-\mv  scaling relation}
	\label{sec-fg}
	For the \mg-\mv scaling relation we obtain $\alpha$ and $\beta$ consistent with zero, i.e. no deviations from self-similarity. The log-normal scatter is $\lesssim 4\%$. The derived self-similar redshift evolution and small intrinsic scatter are in agreement with observational constraints from B19. B19 find a steeper mass dependence ($\alpha \sim 0.26$), while other observational results suggest a mass-slope consistent with self-similarity \citep[e.g.,][]{mantz16}. Note that, these observations are performed at $\Delta_{500c}$. While here we extend to virial radius where the baryon fraction of massive systems, such as those considered here, is approximately the cosmological baryon fraction. Thus the gas fraction ($M_{\rm gas}/M_{\rm vir}$) is approximately constant and independent of the cluster mass i.e. $\alpha = 0$.
	
	Among the cosmological parameters $\gamma$s, we find that the \mg-\mv scaling relation is consistent with being independent of $h_0$, and independent of $\sigma_8$ within $\sim 2 \sigma$. 
	We find \mg $\propto f^{0.8}_b$, which is slightly shallower but significantly away from the dependence expected from a closed box scenario (i.e. \mg $\propto f_b$).
	
\begin{figure}
		\includegraphics[width=9.2cm,angle=0.0]{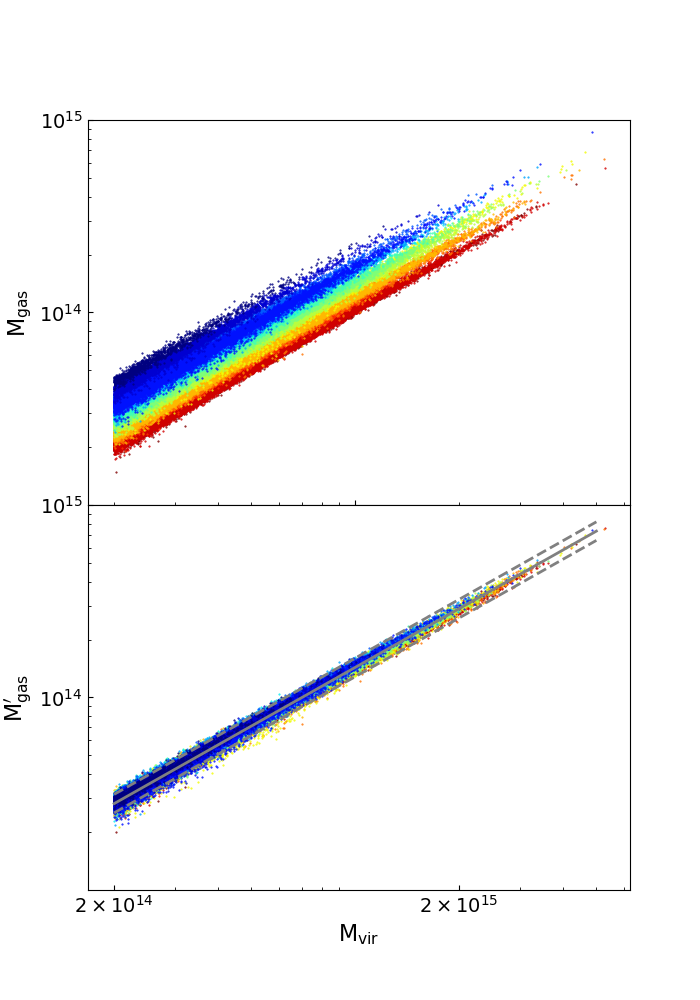}
		\caption{Top panel: Gas mass as a function of halo mass. The color scheme is same as in Figure \ref{fig-cosmo}. 
		Bottom panel: Same as top panel after rescaling all data points to C8 cosmology and absorbing the predicted redshift evolution. The solid and dashed grey lines in the lower panel corresponds to the best-fit and 3$\sigma$ regions at the C8 cosmology.}
		\label{fig-fgas}
	\end{figure}

	In the upper and lower panel of Figure \ref{fig-fgas} we show respectively the original \mg and the predicted $M^\prime_\mathrm{gas}$ (the rescaled \mg at the pivot redshift and C8 pivot cosmology i.e. \mg - redshift dependence - cosmology dependence) as a function of \mv for all the clusters in our sample of all the 15 analyzed cosmologies. The solid and dashed grey lines in the lower panel correspond to the best-fit and 3$\sigma$ regions at the C8 cosmology. We note that the applied rescaling effectively removes the cosmological dependence of the \mg-\mv scaling relation.
	
    \subsection{\mb-\mv scaling relation}	
    \label{sec-mbar}
    For the \mb-\mv relation we find both $\alpha$ and $\beta$ close to zero and hence consistent with the self-similar scenario. The redshift evolution we find is therefore in agreement with current observational constraints \citep{chiu2018}, while we find a mass-dependence slope shallower than \cite{chiu2018}. The log-normal scatter is also remarkably small ($\lesssim 3\%$). 
    We stress however that the observational constraints from \cite{chiu2018} have been derived within different cluster-radii and for cluster samples in a different mass and redshift range, and therefore a direct comparison is not straightforward. 
    
    For the cosmological MOR parameters $\gamma$s, we find both $\gamma_{h_0}$ and  $\gamma_{\sigma_8}$ consistent with zero, i.e. the \mb-\mv relation does not significantly dependent on $h_0$ and $\sigma_8$. We find \mb $\propto f^{0.87}_b$, close to, but still significantly different, than one (the expected closed-box scenario). More in detail, the total baryonic mass has a closer behaviour to the expected closed-box scenario than each individual \mg and \ms component (section \ref{sec-ms}), consistent with the expected anti-correlation of \mg and \ms at fixed halo mass \citep{wu15}. A more focused analysis on the correlation coefficients of different observables at fixed halo-mass will be presented in future works.
    
\subsection{\ms-\mv scaling relation}
\label{sec-ms}
\ms is the sum of mass of all star particles within a given overdensity radius. In the right-hand panel of Fig \ref{fig-uncertain-masses}, we show the cosmology dependence of $\alpha$, $\beta$ and $\sigma$ for \ms-\mv scaling relation (empty triangles). The dark grey shaded region is represents observational uncertainties taken from \cite{chiu2018}, centered at C8 cosmology (see their Table 3). \cite{chiu2018} use DES and {WISE/\it Spitzer} data to constrain \ms-$M_{500c}$ scaling relation for galaxy clusters in the redshift range $0.2 < z < 1.25$, with masses $M_{500c} \gtrsim 2.5\times 10^{14} M_{\odot}$. They assume a simple power law in 1+z as a redshift dependence.

We find that $\alpha$ remains nearly independent of cosmology with $\alpha \sim 0$, close to the self similar prediction (except at C13 where it shows a small deviation from the self-similarity). However, both, the redshift dependence, $\beta$ and the log-normal scatter, $\sigma$ show a strong cosmology dependence. The most plausible explanation for the strong cosmology dependence seen in $\beta$ is that our simplistic assumption, $F(z)= 1+z$ does not capture the redshift evolution of \ms-\mv scaling relation. 

The evolution of the stellar mass fraction of a galaxy is more directly linked to its stellar age rather than its redshift. Each star particle in the {\tt Magneticum} simulations is described by a single stellar population (SSP) model, generated at a given redshift (expansion factor) and then passively evolved. From the post-processing analysis, we first define the age of the galaxy for each sub-halo, corresponding to the average expansion factor of all the associated star particles.
We then average (weighted by the stellar mass of the galaxy) over the expansion factor of all galaxies residing in each main halo to obtain an average redshift of the formation of the stellar content of the cluster. At a given snapshot, the difference between the age of the Universe at that snapshot and the age of the Universe at the formation time of the stellar content represents the stellar age of the halo, $t_*(z)$.

\begin{figure}
	\includegraphics[width=8.5cm,angle=0.0]{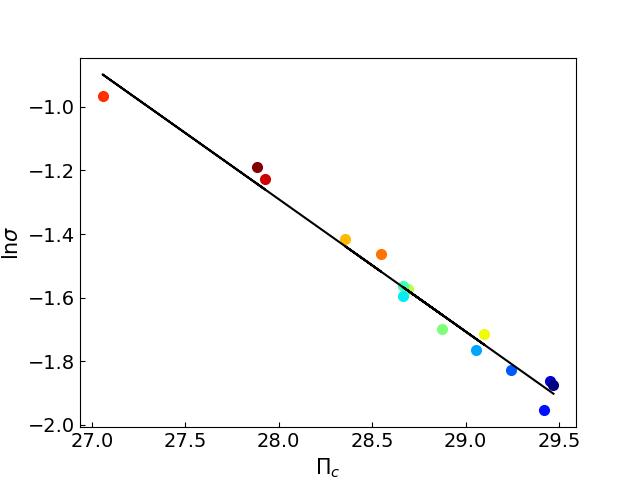}
	\caption{Correlation between scatter and log-normalization of stellar mass scaling relation. Solid-black line follows Equation \ref{eqn-sig-pi}.}
	\label{fig-pi-sigma}
\end{figure}

Using $t_*(z)$ (shown by empty plus markers) instead of $1+z$ (empty triangles) removes most of the cosmology dependence present in $\beta$ as shown in Fig \ref{fig-uncertain-masses}, thus supporting the idea that the time evolution of the galaxy population is a better description. 
However, this parametrization does not help with the cosmology dependence of the scatter $\sigma$. In Fig \ref{fig-pi-sigma}, we show $\sigma$ as a function of $\Pi_c$, i.e. the log normalization of the scaling relation. The black-solid line in the figure follows, 

\begin{equation}
\sigma = \sigma_0 \exp \Bigl[-0.42(\Pi_c- \Pi_{c,0})\Bigr]
\label{eqn-sig-pi}
\end{equation}

where, $\sigma_0 = 0.21$ and $\Pi_{c,0} = 28.69$. This figure suggests that the normalization and the scatter in \ms-\mv scaling relation are tightly correlated, and both of them have similar cosmology dependence. Therefore, instead of further expanding the scatter $\sigma$ as a complicated function of cosmological parameters, we choose to write it a simple function of $\Pi_{c}$. As shown in Fig \ref{fig-uncertain-masses} (filled circles), the combination of $t_*(z)$ redshift evolution and $\sigma(\Pi_c)$ gives us the desired form of \ms-\mv scaling relation.

Consistently with \cite{chiu2018}, we find that the mass slope is smaller than one (i.e. $\alpha+\alpha_{ss} \sim 0.9$). A direct comparison with the redshift evolution is limited by the different adopted functional form and observed overdensities, but it is in general consistent with a mild evolution, at least within the probed redshift range. 
%Coming to the best-fitting values of the scaling relation parameters, we find $\alpha$ is close to zero, i.e. \ms $\propto$ \mv. The best-fitting value of $\beta$ is also consistent with zero. However, the large systematic uncertainties in both $\alpha$ and $\beta$ which are driven by a single cosmology, C13. Excluding C13, we have the redshift evolution roughly consistent with \ms $\propto t^{-0.21} _*(z)$. 
Within the adopted parametrization of the scatter, we find a residual log-normal scatter $\sigma$ close to 20\%. With respect to the cosmological parameters, we find $\gamma_{h_0}$ consistent with zero given the large systematic uncertainty. On the other hand, we find \ms $\propto f^{2.56}_b$ and $\sigma^{1.77} _8$ (with large uncertainty on $\gamma_{\sigma_8}$). This is not too surprising, since star formation is regulated by much more complex physical processes than, for example, the physics describing the relationship between mass and velocity dispersion. Stellar mass is the outcome of physics acting at small scales, such star formation and feedback. At the same time, it is directly affected by large scale cosmology dependent processes, such as cosmological infall and mergers. Therefore, the cosmology dependence of \ms can be strong and non-trivial as suggested by our results.

\subsection{\tg-\mv  scaling relation}
	\label{sec-tg}
	%One of the most well studied and regularly used MOR is the temperature-mass scaling relation. At cluster mass scales, gravity dominates any other physical process shaping the properties of the cluster gas and the ICM is heated to high temperatures ($\sim O(10^7)$ K) through gravitational shock heating. These systems remain close to the hydrostatic equilibrium with \tg $\propto$ $M_{\rm vir}^{2/3}$ $F(z)^{2/3}$. Therefore, the ICM temperature has been used for a long time as a good proxy for the virial mass of the cluster. Recent observations \citep{bulbul18} find a reasonable agreement with the self-similar prediction in the mass dependence of $T_{500}-M_{500}$ scaling relation (within 2-$\sigma$).
	For the \tg-\mv relation, we find a good agreement with the self-similar expectation in the mass slope (e.g., $\alpha \sim 0$), consistent with many observational studies \citep{vikhlinin09, arnaud05, mantz16}. However note that, B19 find a steeper mass slope compared to the self-similar prediction.  
	We find a small negative redshift evolution ($\beta \sim -0.16$) consistent within 1-$\sigma$ with observed redshift evolution (eg. \citealt{mantz16}, B19). We find a log-normal scatter of 11\%, consistent (within 2-$\sigma$) with the observed scatter in \tg-\m5c relation \citep{bulbul18}.
	
	For the cosmology dependence of the scaling relation, we find \tg $\propto h^{0.78} _0 f^{-0.02} _b \sigma^{0.14} _8$. Therefore, the temperature-mass scaling relation is almost independent of the baryon fraction and $\sigma_8$. \tg has a significant dependence on $h_0$, consistent with  the theoretical expectation of 2/3 if \tg $\propto \rho^{1/3} _c$ (and $\rho_c \propto h^2 _0$). The $h_0$ dependencies of \tg and \sigv can also be described by their dependencies on the halo mass, whereas, it is not the case for other observables (see Table \ref{tab-bf-gammas}). However, the $h_0$ dependencies of most of the observables are well described by their dependencies on $\rho_c$. We further highlight that the cosmology dependence of MOR normalization i.e. $\gamma$-parameters represent more an "effective cosmology dependence" and could incorporate the cosmological dependence of other quantities.
		
	\subsection{\yx-\mv scaling relation}
	\label{sec-yx}
	The X-ray integrated pseudo-pressure \yx is an observable which is characterized by a relatively low intrinsic scatter and its closely connected to the SZE observable (\citealt{kravtsov06a, nagai06, nagai07, bonamente08, vikhlinin09, anderson11, benson13, mantz15}; B19).
	The self-similar scenario predicts \yx $\propto M_{\rm vir}^{5/3}$ $F(z)^{2/3}$. We find $\alpha \sim 0$, consistent with the observations by \cite{vikhlinin09, lovisari15, mantz16}, whereas it is shallower than the observations by \cite{arnaud07} and B19. We find $\beta \sim -0.15$, i.e. a small deviation from self-similarity in the redshift dependence as expected from the results for \tg-\mv relation. Observed redshift evolution of $Y$-mass relation is consistent with zero given the large uncertainties (eg. \citealt{mantz16}; B19). The scatter in \yx-\mv relation is around 13\%, driven by the scatter in \tg-\mv relation, and consistent with the observed scatter in \yx-\m5c relation \citep{bulbul18}.
	
	In case of cosmology dependence we find, \yx $\propto f^{0.78} _b$, driven by the strong baryon fraction dependence of gas mass, \yx $\propto h^{0.65} _0$, driven by its temperature dependence and no dependence on $\sigma_8$.
				
	\subsection{\sigv-\mv  scaling relation}
	\label{sec-sigv}
	For perfectly virialized objects, velocity dispersion is tracing the total halo mass, since it is shaped by gravity only. Previous simulations studies (eg. \citealt{evrard08, saro13, munari13}) have already shown that three-dimensional \sigv-\mv relation stays close to the self-similar prediction. We also find $\alpha$ and $\beta \sim 0$. Consistent with previous results, we find a remarkably small scatter in the scaling relation ($\sim$ 5\%). Previous studies have shown, however, that the one-dimensional velocity dispersion-mass relation has a significantly larger scatter, due to halo triaxiality \citep{white10,saro13}.
	
	Coming to the cosmology dependence, we find \sigv$\propto h^{0.38}_0$ and to be independent of baryon fraction and $\sigma_8$, again in agreement with a self-similar scenario (since $\sigma_v^2 \propto \rho^{1/3} _c$).			
				
	\begin{table*}
	\caption{Priors used in our analysis while forecasting the impact of our MOR parameterization on an idealized eROSITA-like experiment (see Section \ref{sec-impact} for details). $\mathcal{U}(a,b)$ represents uninformative prior in the range $(a,b)$. $\mathcal{N}(\mu,\sigma)$ represents Gaussian prior with mean $\mu$ and width $\sigma$. The priors on $\Omega_m$, $h_0$, $\Omega_b h^2 _0$ and $\ln A_s$ are same for all three cases (therefore not shown in table) are given in Section \ref{sec-impact}.}
	\centering
	\resizebox{1.0\textwidth}{!}{
		\setlength{\tabcolsep}{4pt}
		\begin{tabular}{c c c c c c c c}
			\\
			&$\Pi_{c}$&$\alpha$&$\beta$&$\sigma$&$\gamma_{h_0}$&$\gamma_{b}$&$\gamma_{\sigma_8}$\\\\
			\hline \\
			case-(i) & $\mathcal{N}(31.40,0.157)$ & $\mathcal{N}(0.02,0.13)$ & $\mathcal{N}(0.01,0.192)$ & $\mathcal{N}(0.10,0.05)$ & - & - & - \\\\
			case-(ii) & " & " & " & " & $\mathcal{N}(-0.13,0.22)$ &$\mathcal{N}(0.80,0.03)$ & $\mathcal{N}(-0.14,0.06)$ \\\\
			case-(iii) & $\mathcal{N}(31.40 - \ln(10^{\Delta(\Omega_m)},0.157$) & " & " & " & " & " & "\\\\
			\hline
		\end{tabular}
		\label{tab-priors}}
    \end{table*}

	\section{Implication for cosmological studies}
	\label{sec-impact}
	In this section, we forecast the impact of the results presented in the previous section for the parameterization of the MOR on an idealized cluster number-counts cosmology experiment. 
	The simulated experiment resembles a simplified eROSITA
	cluster cosmology analysis, with an idealized gas-mass selected cluster catalog over 15,000 deg$^2$. 
	This analysis does not capture all the sophisticated modelling of the eROSITA selection function \citep[e.g., ][]{grandis18}, but has solely the purpose of highlighting the impact of different cosmological parametrization of the MOR for cluster-cosmology experiments. 
	
	The mock catalog is generated using the \cite{tinker08} halo-mass function assuming a WMAP7 cosmology \citep{komatsu11}. Gas-masses are then computed from the total cluster mass (including intrinsic scatter) using equations \ref{eqn-obs}, and \ref{eqn-pic} and Tables \ref{tab-bf}, and \ref{tab-bf-gammas}. The final catalog consists in all clusters with final $\log (M_{\rm gas}/M_\odot) > 13.75$ (corresponding to a halo mass $\sim 2\times 10^{14} M_{\odot}$) between redshift range 0.1-1.

    We then analyze the above sample, by computing the likelihood of observing the number of clusters $N_{i,j}$ in the survey area for a given redshift bin $i$ and \mg	bin $j$ 
    $\mathcal{L}(N_{i,j}|\vec \theta)$, where $\vec \theta$ contains both the MOR and the cosmological parameters. 
    % We assume the likelihood to be Gaussian and we account in the covariance matrix for the Poisson and sample variance noise.
    In details, we assume a Gaussian likelihood of the form:
    \begin{equation}
    \label{eqn:like}
     \mathcal{L}(d|\theta) \propto \frac{ \exp \left[ -\frac{1}{2} \left(d-m({\vec \theta}) \right)^T {C}^{-1} \left(d-m(\vec \theta) \right) \right]}{\sqrt{(2 \pi)^M {\rm det}(C)}} \, .
    \end{equation}
    where $M$ is the dimensionality of the data vector (three redshift bins $\times$ five gas mass bins = 15), $C$ is the covariance matrix, and $d$ and $m(\vec \theta)$ are respectively the number counts data vector and the corresponding expectation values for the set of parameters $\vec \theta$. The covariance matrix is defined as the sum of a Poisson and sample variance contribution and it is computed analytically at each step of the chain. The two components of the covariance matrix are,
    \begin{equation}
     C^{\rm Poisson} = \delta_{\rm ii} \langle {N} \rangle_i
    \end{equation}
    and
    
    \begin{equation}
    C_{\rm ij}^{\rm Samp\, Var} = \langle b {N} \rangle_i \langle b {N} \rangle_j \sigma^2(V_i,V_j) \, .
    \end{equation}
    where $\langle N \rangle_i$ is the expected number of cluster in the $i$-th bin, while $b$ is the linear halo bias \citep{tinker10}. The last term of equation corresponds to the rms variance of the linear density field within the comoving volume $V_i$, which we approximate with a top-hat window symmetric around the azimuthal axis. We refer the reader to \cite{costanzi18a} for further details on the likelihood and covariance matrix calculation.
	
	The parameter space $\vec \theta$ is explored with the affine invariant sampler {\tt emcee} \citep{emcee}. 
    
    More in detail, the parameters varied during the MCMC analysis are $\Pi_{c,0}$, $\alpha$, $\beta$, $\sigma$ and the cosmological parameters (within a flat $\Lambda$CDM model) $\Omega_m$, $h_0$, $\Omega_b h^2$ and $\ln A_s$. We apply Gaussian priors on $h_0$ ($0.7 \pm 0.05$), $\Omega_b h^2$ ($0.02208 \pm 0.00052$), and uninformative priors on $\Omega_m$ (0.05, 0.5), and $\ln A_s$ (1.0, 6.0). We also apply Gaussian priors on $\Pi_c$ ($31.40 \pm 0.157$), $\alpha$ ($0.02 \pm 0.130$) and $\beta$ ($0.01 \pm 0.192$), where the mean values of the Gaussian are taken from our Table \ref{tab-bf}, and uncertainties are from the estimated uncertainties on the $L_x-M$ from \cite{grandis18} (rescaled appropriately in case of $\alpha$).  
	We apply a Gaussian normal prior on the intrinsic scatter $\sigma_{\ln M_{\rm gas}}$ = $0.10 \pm 0.05$ (the scatter parameter includes the intrinsic and observational scatter) based on the constraint on the scatter of the gas mass-halo mass relation from \cite{mantz16}.
	
	We explore three different scenarios for a cosmological analysis. In all this three cases, we use the above mentioned priors, summarized in Table \ref{tab-priors}.
	
\begin{figure*}
		\includegraphics[width= 0.75 \textwidth]{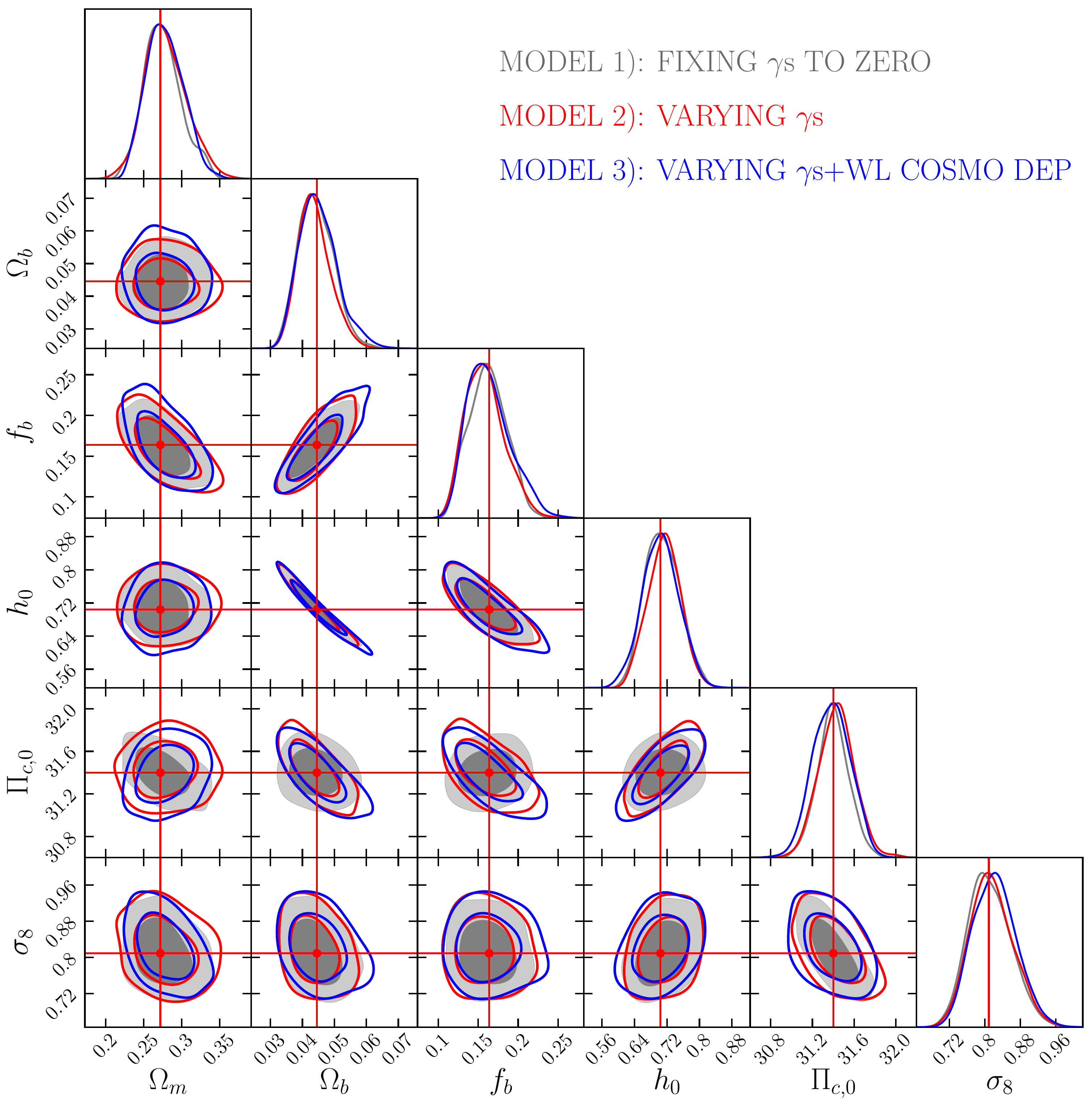}
		\caption{ The 68\% and 95\% confidence limit contours for the scaling relation and cosmological parameters. In grey are the constraints for $\gamma$s fixed to zero, in red the constraints including $\gamma$s in the analysis, and in blue the constraints including $\gamma$s and weak-lensing mass calibration to model the prior on the amplitude of the MOR.}
		\label{fig-gammas}
	\end{figure*}

	\begin{enumerate}
	    \item Fixing $\gamma$'s to zero: 
	    This analysis describes the case where the normalization of the MOR is assumed to be cosmology independent, which represents the approach typically adopted in the literature. 
	    In this case, the amplitude of MOR is simply $\Pi_c = \Pi_{c,0}$. We recover the input parameters in an unbiased way as shown by the grey contours in Fig \ref{fig-gammas}. 
	 
	 \item Varying $\gamma$'s:
	In this case, we also include the cosmological dependence on the MOR, as discussed in Section \ref{sec-method}. We now also include $\gamma$'s as free parameters, with associated Gaussian prior with mean and width taken from Table \ref{tab-bf-gammas}. As a result, the marginalized posterior distributions exhibit now a stronger degeneracy between $\Pi_{c,0}$ and cosmological parameters as shown by red contours in Fig \ref{fig-gammas}. For example, now $\Pi_{c,0}$ shows a strong anti-correlation with $f_b$, driven by the positive value of $\gamma_b$ without any significant variation in the one-dimensional posteriors (as expected due to the uninformative prior on $\Pi_{c,0}$). Such degeneracies otherwise cannot be tested with observational data. Note that the correlation between $\Pi_{c,0}$ and $h_0$ is induced by the tight Gaussian prior applied on $\Omega_b h^2 _0$ and the anti-correlation between $\Pi_{c,0}$ and $f_b$.
	
	\item Varying $\gamma$'s + weak lensing cosmological dependencies on $\Pi_{c}$: 
	In the previous case, we obtained the same marginalized posterior distributions on the cosmological parameters, as expected. The combination of multiple cosmological observables has the potential of breaking degeneracies and provide tighter constraints. Therefore, we now explore the possibility of combining different observables with different sensitivity to the cosmological parameters through their MOR. 
	
	In a real scenario, the amplitude of the gas-mass relation is calibrated with weak-lensing mass estimates. However, weak-lensing calibration also has a cosmological dependence. \cite{simet17} derived matter density dependent mass-richness scaling relation for redMaPPer cluster catalog using weak-lensing data from SDSS. Within their adopted model, the amplitude of the scaling relation depends only on the assumed value of $\Omega_M$ (in a range $\sim$ 0.26-0.34) with a linear decline of log-amplitude as $\Omega_M$ increases. We use their weak-lensing mass calibration to model the prior of our MOR. The amplitude of $M_{\rm gas}-M_{\rm halo}$ relation can now be written as,
	\begin{equation}
	    \Pi_c (\Omega_m) = \Pi_c - \ln{10^{\Delta (\Omega_m)}}
	\end{equation}
	where, $\Delta(\Omega_m) = \alpha  \frac{d \log M_{\rm WL}}{d\Omega_m} (\Omega_M -0.3)$. Incorporating above dependencies modifies the mean of the Gaussian prior applied on $\Pi_{c}$ to $31.40 - \ln{10^{\Delta (\Omega_m)}}$, whereas, the width of the Gaussian remains unchanged. We measure the gas-mass and the weak lensing mass, assuming a reference cosmology and then calibrate the normalization of the gas-mass relation at that reference cosmology. We then explore the cosmological parameter space at each new cosmology. We rescale the gas-mass relation (assuming our MOR), and the constraints we obtain from weak lensing as if we were at the new cosmology. The results of the analysis are shown by blue contours in Fig \ref{fig-gammas}. Addition of priors from weak-lensing experiments does not impact most of the model parameters due to the different cosmological dependencies of the weak-lensing (which is mostly sensitive to $\Omega_m$) and $M_{\rm gas}-M_{\rm vir}$ scaling relation (which is proportional to the baryon fraction). However, there is a slight improvement in the constraints on $\Omega_m$ (from $0.279^{+0.059}_{-0.051}$ in the previous case to $0.278^{+0.046}_{-0.046}$).

    To compare the performance of the above described models, we use Deviance Information Criterion (DIC) and interpret the results using Jeffrey's scale. For a given model $M_i$, DIC is defined as \citep{spiegelhalter02},
\begin{equation}
    {\rm DIC}(M_i) = \langle \chi^2 \rangle + p_d,
\end{equation}
where $\langle \chi^2 \rangle = -2 \langle \ln \mathcal{L}(d|\theta,M_i) \rangle$ is the mean of $\chi^2(\theta)$ and the term $p_d$ is called Bayesian complexity defined as, $p_d = \langle \chi^2 \rangle - \chi^2(\Tilde{\theta})$, where $\Tilde{\theta}$ is the maximum likelihood (minimum $\chi^2$) point. A model which describes the data vector better has a lower value $p_d$ and a higher likelihood and thus a lower value of DIC \citep{grandis16}. 

For the three scenarios considered in this paper we find, DIC$(M_2)$ - DIC$(M_1) = -4.1$, DIC$(M_3)$ - DIC$(M_1) = -4.5$ and DIC$(M_3)$ - DIC$(M_2) = -0.4$. Therefore, both the models with non-zero values of $\gamma$s (Model 2 and Model 3) have a positive preference over Model 1. However, our final model Model 3 including priors from WL studies has an insignificant preference over model Model 2.
    
    To summarize, our Model 1 represents a typical approach where the amplitude of the scaling relation is assumed to be independent of the cosmological parameters. Model 2 is based on our suggested MOR parameterization, which we find a more appropriate choice as it now includes correlations between the MOR normalization and cosmological parameters and therefore better represents our derived theoretical priors. We stress that, by design, the parametrization adopted in Model 2 does not weaken the constraining power of the data on the marginalized posteriors. Finally, Model 3 represents an idealized case to demonstrate how the combination of different observables (which are subject to different degeneracies) can exploit the theoretical priors we derived. This scenario has thus the potentiality of breaking these degeneracies in order to provide tighter cosmological constraints.
	\end{enumerate}
	
	\section{Summary}
	\label{sec-summary}
	{\tt Magneticum} simulations provide one of the most powerful tools for exploring large scales in the presence of both gravitational and complex hydro-dynamical processes in different cosmological environments. In this paper, we studied the mass-observable scaling relation for clusters ($M_{\rm vir} > 2 \times 10^{14} M_{\odot}$) up to redshift $z < 1$. We examined the  cosmology dependence of the $M_{\rm gas}-M_{\rm vir}$, $T_{\rm gas}-M_{\rm vir}$, $\sigma_v-M_{\rm vir}$ and $Y-M_{\rm vir}$ relations using fifteen large cosmological boxes produced with the {\tt Magneticum} simulation set-up with varying cosmological parameters. These cosmologies are chosen using Latin hypercube sampling to fairly sample the constraints on $h_0$, $\sigma_8$ and $\Omega_m$, obtained from the latest SPT cluster number count results. 
	We divide the MOR parameters into two categories: $i$) the astrophysical ones, describing the normalization ($\Pi_{c,0}$), the mass-slope ($\alpha$), the redshift evolution ($\beta$), and the intrinsic lognormal scatter ($\sigma$), and $ii$) the cosmological ones ($\gamma_{h_0}$, $\gamma_{b}$ and $\gamma_{\sigma_8}$), describing respectively the impact of $h_0$, the baryon fraction $f_b$ and $\sigma_8$, on the amplitude of the scaling relations. 
	
	All four observables considered here show a perfect self-similar mass dependence i.e. $\alpha$ consistent with zero. The redshift dependence of \mg and \sigv are in good agreement with the self-similar prediction whereas \tg and \yx show a small deviation. The scatter in $M_{\rm gas}-M_{\rm vir}$ relation is smallest (3-4\%) among the four observables followed by \sigv ($\sim 5\%$), \tg($\sim 11\%$) and \yx ($\sim 13\%$). We investigate the cosmology dependence of $\alpha$, $\beta$ and $\sigma$, and do not find any significant variation, a result that therefore justifies our assumed functional form for the adopted scaling relations.
	
	With respect to the cosmological parameters, we find that the $h_0$ dependence of the MOR agrees with the theoretical expectation, where the scaling of the observables is associated with the variation of the critical density of the Universe (as a function of both redshift and cosmology). 
	
	We find $M_{\rm gas} \propto f^{0.8} _b$, i.e. the gas mass - halo mass scaling relation is slightly shallower but significantly away from the value expected from a closed box scenario ($M_{\rm gas} \propto f_b$) given the uncertainty in $\gamma_b$. $T_{\rm gas}$ and $\sigma_v$ are instead found to be independent of $f_b$, whereas, $Y \propto f^{0.78} _b$ driven by the baryon fraction dependence of the gas mass. 
	
	The cosmological dependence on $\sigma_8$ is consistent with zero for all the four studied observables.
	
	In order to provide robust uncertainties on the MOR parameters (more reliable than the negligible statistical ones), we estimate the systematic uncertainty of the MOR by propagating the error associated with our choice of the functional form used to describe the scaling relations.
	
	As a proof of concept, we show the impact of the cosmological dependence of the MOR for an idealized eROSITA-like cluster cosmology experiment.
	More in detail, we show that our cosmology dependent parametrization introduces a strong degeneracy between the amplitude of the scaling relation and the cosmological parameters, without affecting the one-dimensional marginalized posterior distribution and it is preferred over the cosmology independent normalization model. However, the combination of different observables, which are subject to different cosmological dependencies, can help in breaking these degeneracies and therefore provide a powerful way to tighten cosmological constraints. 
	
	Upcoming next generation surveys will allow us to calibrate the astrophysical parameters of MORs with unprecedented accuracy. While multi-wavelength data-sets are fundamental to directly constrain the astrophysical parameters from observations, the calibration of the cosmology dependence of the MOR is only possible through the analysis of cosmological hydro-dynamical simulations or accurate theoretical modelling of cluster formation and evolution. This work represents therefore a step towards understanding the impact of the cosmological dependence of galaxy cluster scaling relations with hydro-dynamical simulations for future cluster cosmology experiments.\\\\
	{\bf{ACKNOWLEDGEMENTS}}\\
	We thank the anonymous referee for many insightful suggestions and comments. We thank Sebastian Bocquet, Stefano Borgani, August Evrard, Salman Habib, Joseph Mohr, Daisuke Nagai and Elena Rasia for stimulating discussions and their valuable suggestions. PS, MC, AS are supported by the ERC-StG ‘ClustersXCosmo’ grant agreement 716762,  AS is supported by the FARE-MIUR grant 'ClustersXEuclid' R165SBKTMA. Computations have been performed at the ‘Leibniz-Rechenzentrum’ with CPU time assigned to the Project “pr83li” and "pr74d0". KD acknowledges support through ORIGINS, founded by the Deutsche 
    Forschungsgemeinschaft (DFG, German Research Foundation) under Germany's Excellence Strategy – EXC-2094 – 390783311 and by the DAAD, contract number 57396842 .
	
	\footnotesize{
		\bibliography{bibtexcgm,spt}{}
		\bibliographystyle{mn2e}
	}

\begin{appendices}
\renewcommand\thefigure{\thesection.\arabic{figure}}    
\renewcommand{\thetable}{A\arabic{table}}
\renewcommand{\theequation}{A\arabic{equation}}

\section{\lx-\mv scaling relations}
\setcounter{figure}{0}    
\setcounter{table}{0}
\setcounter{equation}{0}
\label{sec-lx}
Bolometric gas luminosity \lx is the sum of emissivity of all gas particles within a given overdensity radius. The self-similar evolution predicts \lx$\propto M_{\rm vir}^{4/3} F(z)^{7/3}$. We find a strong deviation from this prediction as well as a more substantial cosmology dependence of this deviation.

\begin{figure}
	\includegraphics[width=9.0cm,angle=0.0]{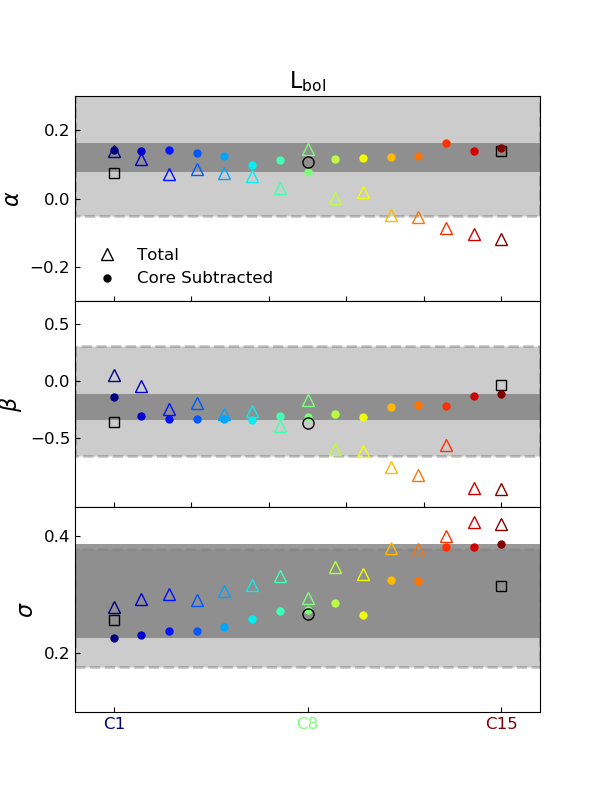}
	\caption{Empty triangles are for total gas luminosity whereas filled circles are for core-subtracted gas luminosity. Rest is the same as Figure \ref{fig-uncertain-masses}.}
	\label{fig-lxms}
\end{figure}

In Figure \ref{fig-lxms}, we show $\alpha$, $\beta$ and $\sigma$ for the  $L_{\rm bol}-M_{\rm vir}$ relation (empty triangles). The dark shaded region represents observational uncertainties taken from \cite{bulbul18}. All three parameters exhibit a large variation as we move from C1 to C15. Most of the X-ray luminosity is associated to the central cluster regions, as $L_{\rm bol}$ is proportional to the density square. For this reason, it is much more strongly affected by relatively small scale physical processes such as cooling and feedback, which are not naturally accounted by the self-similar prediction \citep{pratt09, bulbul18}. 

To confirm this, we replace the total bolometric luminosity by the core-subtracted luminosity, $L^{cs} _{\rm bol}$.
Observers generally remove the contribution coming from central 15\% of $R_{500c}$ to obtain the core-subtracted luminosity.
However, we do not have information stored for individual particles. Instead we have observable quantities for six overdensities, 
$\Delta_i = \rm 2500c, \, 500c, 500m, \, 200c, \, 200m$ and vir, where the subscript c and m correspond to the critical density and mean matter density of the Universe, respectively. We obtain $L^{cs} _{\rm bol}$ by removing the contribution coming from the region within $2500  \rho_c$, closest the traditional definition of core-subtracted luminosity. This removes most of the the cosmology dependence for the resulting mass and redshift slopes $\alpha$ and $\beta$ (shown as filled circles and the systematic uncertainties shown by light grey shaded region enclosed between dashed lines in Figure \ref{fig-lxms}). We find that the associated log-normal scatter for the $L^{cs} _{\rm bol}-M_{\rm vir}$ relation is reduced, however, it still shows some variation with cosmology.

Results of the fitting procedure for the core-subtracted luminosity are shown in Table \ref{tab-bf-apx}. There is a minor deviation from the self-similar mass slope compared to the observed core-subtracted $L_{\rm X,bol}-M_{500c}$ relation \citep{bulbul18}. Note that the observations are made at a different overdensity and a different definition of core radius. The redshift slope is in agreement with observed as well as the self-similar prediction. The scatter in the scaling relation is $\sim 31\%$, in a good agreement with the observed scatter in $L_{\rm X,bol}-M_{500c}$ relation. There are large uncertainties on $\gamma_{h_0}$ and $\gamma_{\sigma_8}$ and both of them consistent with zero within $2-\sigma$. However, $L^{cs} _{\rm bol}-M_{\rm vir}$  exhibit a strong dependence on the baryon fraction with $L^{cs} _{\rm bol} \propto f^{1.93} _b$. This dependence is expected as the gas luminosity is strongly depends on the underlying gas density ($L_{\rm bol} \propto \rho^2$).  

	\begin{table}
	\caption{Same as Table \ref{tab-bf} \& \ref{tab-bf-gammas} for \lx-\mv scaling relation (core-subtracted bolometric luminosity as defined in Section \ref{sec-lx}).}
	\centering
	\resizebox{0.18 \textwidth}{!}{
		\setlength{\tabcolsep}{8pt}
		\begin{tabular}{c c}
			\\
			& \lx \\\\
			\hline \\
			$\Pi_{c,0}$ & -0.43$\pm$0.52 \\\\
			$\alpha$ & 0.12$\pm$0.04 \\\\
			$\beta$ & -0.22$\pm$0.115 \\\\
			$\sigma$ & 0.31$\pm$0.08 \\\\
			\hline \\\\
			$\gamma_{h_0}$ & 1.03$\pm$0.61 \\\\
			$\gamma_{b}$   & 1.93$\pm$0.06 \\\\
			$\gamma_{\sigma_8}$ & -0.37$\pm$0.30 \\\\
			\hline
		\end{tabular}
		\label{tab-bf-apx}}
\end{table}

\section{Scaling relations for \m5c}
\label{sec-m500c}
\setcounter{figure}{0}    
\setcounter{table}{0}
\renewcommand\thefigure{\thesection.\arabic{figure}}    
\renewcommand{\thetable}{B\arabic{table}}

We presented our results for $O-M_{\rm vir}$ relations in the main analysis. In this section, we present our results for $O-M_{\rm 500c}$ relations. Again note that the observables are measured within $R_{\Delta _i}$ when considering the scaling relation $O-M_{\rm {\Delta _i}}$ (except for \sigv which is independent of the over-density definition). Another difference between \mv and \m5c MORs is the self-similar redshift evolution. We replace $F(z)$ by $E(z)$ in Equation \ref{eqn-obs} while dealing with $O-M_{\rm 500c}$ relations. The mass and redshift pivots and the redshift cutoff remain the same. The lower mass cutoff is now at $M^{\rm cutoff} _{500c}=2\times 10^{14}$.  

\begin{figure*}
	\hspace*{-17mm}
	%	\centering
	\includegraphics[width=22.0cm,angle=0.0]{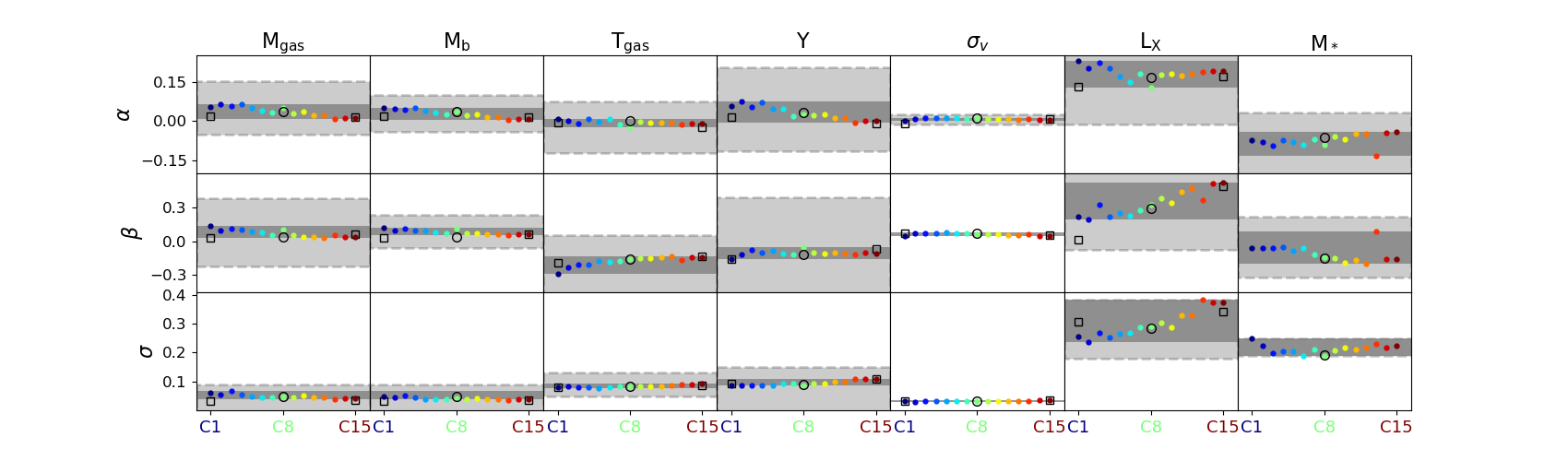}
	\caption{Same as Figure \ref{fig-uncertain-masses} for $M_{500c}$ MORs. For gas luminosity, we show only the results for core-subtracted luminosity and for stellar mass, we show the results for $t_*(z)$ redshift parametrization with scatter as a function of log-normalization of the scaling relation.}
	\label{fig-uncertain-m550c}
\end{figure*}

In Figure \ref{fig-uncertain-m550c}, we show the variation in astrophysical MOR parameters as a function of cosmology and in Table \ref{tab-bf-apx2} we list the best-fitting results and systematic uncertainties for all six observables. 
In general, there is a small increase in the variation of $O-M_{\rm 500c}$ parameters from C1 to C15 compared to that of virial over-density. For \mg, \mb, \tg, \sigv and \yx, the systematic uncertainties are well within the observational uncertainties. In the case of \lx and \ms, systematic uncertainties in the log-normal scatter are comparable to the observed ones, similar to what we found for \mv scaling relations. Given the systematic uncertainties, there are no major differences between the best-fitting results for \mv and $M_{500c}$ (except in the case of the redshift evolution of $L_{\rm bol}$).

\begin{table*}
	\caption{Same as Table \ref{tab-bf}, \ref{tab-bf-gammas} and \ref{tab-bf-apx} for $O-M_{\rm 500c}$ scaling relation.}
	\centering
	\resizebox{0.8 \textwidth}{!}{
		\setlength{\tabcolsep}{8pt}
		\begin{tabular}{c c c c c c c c}
			\\
			& \mg & \mb & \ms & \tg & \yx & \sigv & \lx \\\\
			\hline \\
			$\Pi_{c,0}$ & 31.28$\pm$0.34 & $31.36 \pm 0.38$ & 28.47$\pm$1.19 & 1.06$\pm$0.08 & 32.33$\pm$0.35 & 6.55$\pm$0.02 & 0.13$\pm$0.75\\\\
			$\alpha$ & 0.04$\pm$0.03 & $0.03 \pm 0.02$ & -0.09$\pm$0.04 & -0.01$\pm$0.02 & 0.03$\pm$0.04 & 0.01$\pm$0.01 & 0.18$\pm$0.05 \\\\
			$\beta$ & 0.08$\pm$0.05 & $0.09 \pm 0.03$ & -0.05$\pm$0.14 & -0.22$\pm$0.08 & -0.11$\pm$0.05 & 0.06$\pm$0.01 & 0.36$\pm$0.17 \\\\
			$\sigma$ & 0.05$\pm$0.01 & $0.04 \pm 0.01$ & 0.22$\pm$0.03 & 0.08$\pm$0.01 & 0.10$\pm$0.01 & 0.032$\pm$0.002 & 0.31$\pm$0.07 \\\\
			\hline \\\\
			$\gamma_{h_0}$ & -0.27$\pm$0.27 & $-0.03 \pm 0.22$ & 1.27$\pm$0.67 & 0.81$\pm$0.18 & 0.59$\pm$0.35 & 0.36$\pm$0.01 & 1.45$\pm$0.52 \\\\
			$\gamma_{b}$ & 0.70$\pm$0.04 & $0.82 \pm 0.04$ & 2.57$\pm$0.06 & 0.13$\pm$0.05 & 0.79$\pm$0.06 & -0.015$\pm$0.001 & 1.85$\pm$0.07 \\\\
			$\gamma_{\sigma_8}$ & -0.22$\pm$0.08 & $-0.14 \pm 0.06$ & 1.65$\pm$0.39 & 0.10$\pm$0.05 & -0.10$\pm$0.07 & -0.05$\pm$0.01 & -0.65$\pm$0.24 \\\\
			\hline
		\end{tabular}
		\label{tab-bf-apx2}}
\end{table*}

\section{Variation in subgrid prescription}
\label{sec-subgrid}
One of the major source of uncertainty in our analysis is fueled by the unknown cosmology dependence of subgrid prescription. In the main analysis, the subgrid model parameters are tuned to the observations at C8, and we then do not vary these parameters for other cosmologies. In this section, we discuss the impact of variation in subgrid prescription on the MOR parameters. 

We run non-radiative versions of C1 and C15 cosmologies (shown by empty squares in Figures \ref{fig-uncertain-masses}, \ref{fig-uncertain}, \ref{fig-lxms} \& \ref{fig-uncertain-m550c}). To explore the influence of the sub-grid models we modified the most 
influential parameters defining the strength of the feedback, namely the velocity of the galactic
wind (kinetic feedback) as well as the overall strength of the AGN feedback in simulations. For the 
later we run two simulations where the feedback efficiency of the AGN model was changed from
the default value of 0.15 (as used in the simulations varying the cosmology) to either be 0.1 (a1 run) 
or 0.2 (a2 run). Furthermore, we also explored the influence of the galactic wind feedback, by changing 
our default value of 350 km/s to 500 km/s (w1 run) and 800 km/s (w2 run). While the comparison with 
the non radiative runs give a general feeling of the importance of the detailed star-formation and black hole 
model, this should capture the range of reasonable choices of the details within the according sub-grid
models. We do not find any significant difference in the best-fitting values of the astrophysical parameters when we vary the subgrid model parameters as shown by empty circles in Figures \ref{fig-uncertain-masses}, \ref{fig-uncertain}, \ref{fig-lxms} \& \ref{fig-uncertain-m550c}. For clarity, we plot the results for a2 run only in the figures since the differences between the parameter constraints obtained from different feedback runs are negligible. For most of the observables, the variation is well within the systematic uncertainties and for all of them it is much smaller than the observed uncertainties.

\end{appendices}
	
\end{document}